\documentclass[aps,prc,tightenlines,floatfix,showpacs,twocolumn]{revtex4}

\usepackage{graphics}
\usepackage{bm}
\usepackage{amsmath}
\usepackage{amssymb}
\usepackage{supertabular}
\usepackage{array}

\newcommand{\sfrac}[2]{\textstyle\frac{#1}{#2}}

\begin{document}

\title{Mass-15 nuclei and predicting narrow states beyond the proton drip line.}

\author{P. R. Fraser$^{(1)}$}
\email{paulfr@trinity.unimelb.edu.au}
\author{K. Amos$^{(2,4)}$}
\email{amos@unimelb.edu.au}
\author{L. Canton$^{(3)}$}
\email{luciano.canton@pd.infn.it}
\author{S. Karataglidis$^{(4)}$}
\email{stevenka@uj.ac.za}
\author{D. van der Knijff$^{(2)}$}
\email{dvanderknijff@gmail.com}
\author{J. P. Svenne$^{(5)}$}
\email{svenne@physics.umanitoba.ca}

\affiliation{$^{(1)}$
Trinity College, University of Melbourne, Victoria 3010, Australia}
\affiliation{                    
${}^{(2)}$ School of Physics, University of Melbourne, Victoria 3010,
Australia}
\affiliation{$^{(3)}$ Istituto Nazionale di Fisica Nucleare,
Sezione di Padova, via Marzolo 8, Padova I-35131, Italia,}
\affiliation{$^{(4)}$ Department of Physics, University of
Johannesburg, P.O. Box 524 Auckland Park, 2006, South Africa}
\affiliation{$^{(5)}$ Department  of  Physics  and Astronomy,
University of Manitoba, and Winnipeg Institute for Theoretical
Physics, Winnipeg, Manitoba, Canada R3T 2N2}

\date{\today}

\begin{abstract}
In a previous letter (Phys. Rev. Lett. 96, 072502 (2006)), the
multi-channel algebraic scattering (MCAS) technique was used to
calculate spectral properties for proton-unstable $^{15}$F and its mirror,
$^{15}$C.  MCAS achieved a close match to the then-new data for $p+^{14}$O
elastic scattering and predicted several unusually narrow resonances
at higher energies.

Subsequently, such narrow resonance states were found. New cross
section data has been published characterising the shape of the $J^\pi
=\frac{1}{2}^-$ resonance.  Herein we update that first MCAS analysis
and its predictions.  We also study the spectra of the set of mass-15
isobars, ${}^{15}$C, ${}^{15}$N, ${}^{15}$O, and ${}^{15}$F, using the
MCAS method and seeking a consistent Hamiltonian for clusterisation
with a neutron and a proton, separately, coupled to core nuclei
${}^{14}$C and ${}^{14}$O.
\end{abstract}


\maketitle


\section{Introduction}

The low-energy spectra of exotic, light-mass nuclei beyond the drip
lines have been the foci of intense research efforts since the advent
of radioactive ion beams.  The nucleus $^{15}$F has been of special
interest both as it spontaneously emits a proton, and for the role
played by that reaction in the $2p$-decay of $^{16}$Ne.

Herein we report on results of calculations of the low energy spectra
of the mass-15 isobars, ${}^{15}$C, ${}^{15}$N, ${}^{15}$O, and
${}^{15}$F. These nuclei are disparate in that ${}^{15}$O and
${}^{15}$N have deep binding and many fully bound states in their low
energy spectra, while ${}^{15}$C is weakly bound with only two
subthreshold (to neutron emission) states and ${}^{15}$F is unbound
(to proton emission). To describe the low energy spectra of these
systems with a single, simple, Hamiltonian is the difficult aim we
set.  However, a primary focus under this aim is to predict the
existence and location of more states in the exotic nucleus,
${}^{15}$F, than are currently known.  Regarding this nucleus, in 2002
a $p+^{14}$O Wood-Saxon potential was parameterised~\cite{Gr02} to
find the energies and widths of the only two $^{15}$F states then
known; the ground $J^\pi = \sfrac{1}{2}^+$ and first excited
$\sfrac{5}{2}^+$ resonances.  That potential, used in a three-body
model for ${}^{16}$Ne, proved useful in recent analyses of 2-proton
decay data~\cite{Br14,Gr15}.  In 2004, the first
$^{14}$O($p,p$)$^{14}$O cross section data taken at several angles and
at energies spanning the two known resonance states was
published~\cite{Go04}.  In that paper, data fits found using
Woods-Saxon potentials were shown.  The next year, this data was
analysed with a microscopic cluster model, which obtained a good
match~\cite{Ba05}, and further data was soon taken and
published~\cite{Gu05}. In the same year, properties of these two
$^{15}$F states were studied with a simplistic shell
model~\cite{Fo05}.  This model was restricted to the lowest three
configurations of one-particle/two-hole and three-particle/four-hole, states.

In 2006, the multichannel algebraic scattering method (MCAS) was used
to analyse the data of Refs.~\cite{Go04,Gu05}, defining potentials
between $^{14}$O and protons from a collective model with rotor
character, while accounting for the Pauli principle between the proton
and the underlying $^{14}$O shell structure~\cite{Ca06}. 
As well as obtaining a close fit with
the cross-section data, the MCAS calculation predicted narrow
resonances at higher energies. These were a $\sfrac{1}{2}^-$ state
with energy (width) of 5.49 (0.005) MeV, a $\sfrac{5}{2}^-$ of 6.88
(0.01) MeV, a $\sfrac{3}{2}^-$ of 7.25 (0.04) MeV, as well as
$\sfrac{1}{2}^+$, $\sfrac{5}{2}^+$ and $\sfrac{3}{2}^+$ states of 7.21
(1.2), 7.75 (0.4) and 7.99 (3.6) MeV, respectively.

The width of such narrow states caused some
controversy~\cite{Fo07,Ca07} (with Ref.~\cite{Fo07} using a potential
model to construct broad single-particle resonances whose widths were
manually scaled down by over an order of magnitude to fit data for
narrow resonances). Subsequently however, the existence of the states
predicted by the MCAS calculation has been verified
experimentally~\cite{Mu08,Mu09,Mu10}. (Note that in Table I of
Ref.~\cite{Mu10}, the labels for results reproduced from
Ref.~\cite{Ca06} and \cite{Fo07} were accidentally switched.)  For
completeness we note that the afore-mentioned simple shell model
calculation was revised~\cite{Fo11} in the light of the new data.

Narrow states have now been observed in other proton rich nuclei, e.g.
$^{19}$Na~\cite{Pe08}, $^{16}$Ne~\cite{Mu10}, $^{15}$Ne~\cite{Wa12},
and $^{23}$Al~\cite{Ga08}, with narrow resonances of the latter found
with an MCAS study~\cite{Fr16}.  They have been predicted for
$^{21}$Al~\cite{Ti12} and $^{25}$P~\cite{Fe15}.
Such narrow resonances indicate an eigenstate with structure which has
little overlap with the ground state. In the case at hand, this is the
difference between a one-proton emitting clusterization ($p+^{14}$O) and a
two-proton emitting clusterization ($2p+^{13}$N). Pauli hindrance accounts
for this effect~\cite{La15}.

Recent developments include the publication of more complete data with
smaller uncertainties over a larger energy
range~\cite{Ol11,Ol11a,Gr16}. Where Ref.~\cite{Mu10} provided evidence
of the narrow $\frac{1}{2}^-$ resonance predicted by MCAS,
Ref.~\cite{Gr16} provides details of its shape, finding it to be a dip
in cross section, as did our first MCAS calculation. Further, in
Ref.~\cite{Gr16} and in a recently-published thesis~\cite{Me16}, the
coupled-channels Gamow shell model (GSM-CC)~\cite{Ja14} has been used
to calculate the $^{14}$O($p,p$)$^{14}$O cross section in the energy
range of that data, reproducing the $\frac{1}{2}^+$, $\frac{5}{2}^+$
and $\frac{1}{2}^-$ resonances well.  At higher energies, that
calculation slightly underestimates experiment.

As MCAS theory has undergone a decade of refinement since
Ref.~\cite{Ca06}, we now take the opportunity presented by this new
data to revisit our calculation in the energy range where cross
sections have been measured and beyond, where MCAS predicts further
resonances. Section~\ref{method} summarises the MCAS method and
details improvements since the work of Ref.~\cite{Ca06}.
Section~\ref{spectra} presents calculated results for the spectra of
the mirror systems ${}^{15}$C and ${}^{15}$F. Section~\ref{xsect}
shows the new $p+^{14}$O cross section results compared to recent
data. In Section~\ref{ON}, we investigate how many details of the
spectra of another mass-15 mirror pair, $^{15}$O and $^{15}$N, may be
described by essentially the same nuclear potential, i.e. that for
$n$+$^{14}$O and $p$+$^{14}$C. Finally, conclusions are drawn in
Section~\ref{conc}.

\section{Details of the method}
\label{method}

The method finds solutions of coupled-channel Lippmann-Schwinger
equations in momentum space using finite-rank expansions of an input
matrix of nucleon-nucleus interactions. 
A set of Sturmian functions is used as expansion basis and this allows
locating all compound-system resonance centroids and widths,
regardless of how narrow, and by using negative energies, allows
determination of sub-threshold bound states.  Further, use of
orthogonalizing pseudopotentials (OPP) in generating the Sturmians
ensures that the Pauli principle is not violated~\cite{Ca05,Am13},
even with a collective-model formulation of nucleon-nucleus
interactions. Otherwise, some compound nucleus wave functions may
possess spurious components~\cite{Am05}.

Results we have obtained vary slightly from those previously
published~\cite{Ca06} since five target (or core) nuclear states now
have been used in the coupled-channel evaluations (rather than the
three in Ref.~\cite{Ca06}), and so the interaction potential
parameters have been adjusted slightly, and exact masses of the
nucleons and nuclei used rather than the mass numbers. Further,
the Coulomb interactions in the $p$+${}^{14}$O cluster has been
derived from a three parameter Fermi (3pF) form for the charge distribution in ${}^{14}$O,
adding to the nuclear interaction which has the form
\begin{multline}
V_{cc'}(r) = \bigg[ \left\{ V_0 + V_{ll} {\mathbf \ell\cdot \ell} \right\} 
f(r,R,a) 
\\ \left. + V_{\ell\cdot s} \frac{df(r,R,a)}{dr} 
{\mathbf \ell\cdot s}\right]_{cc'} .
\label{interact}
\end{multline}
Here $f(r,R,a)$ is a deformed Woods-Saxon function, and both
quadrupole and octupole deformations are taken
to second order in specifying the coupled-channel ($c,c'$) potentials.

For full details, see ~\cite{Ca05,Am03,Fr08a}.  

\subsection{States used for the core nuclei, {\rm $^{14}$C} and {\rm$^{14}$O}}

\begin{figure}[h]
\scalebox{0.53}{\includegraphics*{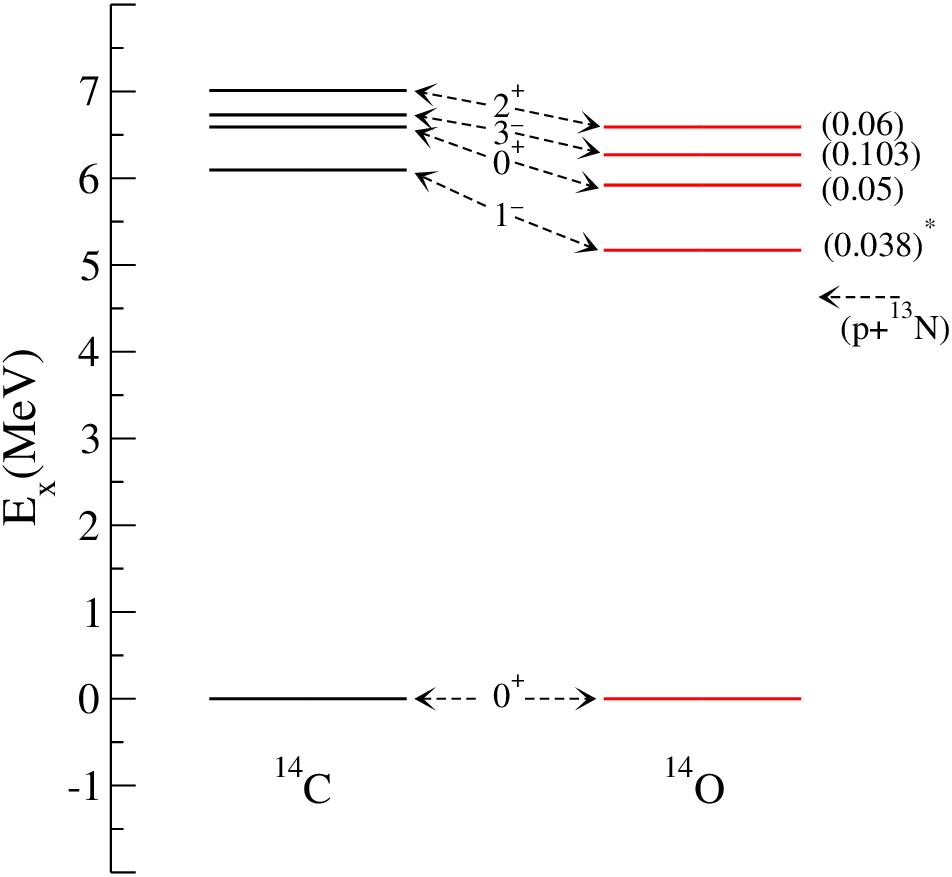}}
\caption{\label{Mass-14-lev}
The low excitation spectra of the mass-14 mirror nuclei,
${}^{14}$C and ${}^{14}$O, used
in MCAS calculations. The spin-parities of the states are listed
in the middle of the diagram.
}
\end{figure}

In Fig.~\ref{Mass-14-lev}, the known low-energy spectra of the mirror
nuclei ${}^{14}$C and ${}^{14}$O are shown.  These states have all
been used in the current coupled-channel calculations.  While the
sequence of each of the states shown (the spin-parities) are as
required by the mirror condition and the excitation energies are
comparable, there are features that vary from a strict mirror
arrangement.  Notably, the actual excitation energies of the states in
${}^{14}$O differ from those of their matching partners in ${}^{14}$C,
as do the energy gaps, but also the relative nucleon breakup energies
are quite different; 8.176 MeV for neutron emission from ${}^{14}$C
but only 4.628 MeV for a proton emission from ${}^{14}$O. Consequently
the four excited states in ${}^{14}$O are resonances while those in
${}^{14}$C are not.  The widths of the four ${}^{14}$O resonances are
shown in brackets in Fig.~\ref{Mass-14-lev} and the units are MeV. The
asterisk with the width of the first excited ($1^-$) resonance
indicates that its emission form is not identified in the tabulation
used~\cite{Aj91}. The other three all decay by proton emission. For
details of how MCAS treats core nuclei states which are themselves
resonances, see Refs.~\cite{Fr08a,Ca11,Fr16a}.  This represents
another upgrade with respect to the calculation originally
published~\cite{Ca06}.

As the two cores used in these coupled-channel,
nucleon-nucleus cluster calculations do not show perfect mirror
symmetry even at low excitation, one may expect the possibility of
some asymmetry between the two Hamiltonians required to best define
the relative mass-15 spectra, in addition to simply a Coulomb
interaction added to the cluster model Hamiltonian that best
describes the ${}^{15}$C spectrum.  Some added asymmetry may be due to a
charge dependence of the strong force.  With the $NN$ interaction,
evidence for charge-symmetry and charge-independence breaking is given
by the results of scattering experiments; asymmetry has been noted
with scattering lengths, namely
\begin{align}
&a^N_{pp} - a^N_{nn} = 1.65\pm 0.60\ ,\\
{\rm and}\hspace*{1.0cm}&
\nonumber\\
&(a^N_{pp} + a^N_{nn})/2 -a^N_{np} = 5.6\pm 0.6 .
\end{align}
The first indicates a small difference between $v^N_{pp}$ and
$v^N_{nn}$ implying a charge-symmetry breaking, while the second, is
evidence of the breaking of charge-independence.  So $NN$ forces have
a charge dependence and that suggests there may be a non-negligible
isospin-symmetry breaking component of the effective $NN$ interaction
to be used in models of nuclear structure. The isospin non-conserving
(INC) shell model~\cite{La13} is an example.

When considering how much charge dependence may affect differences
between the two mirror nuclei considered, it is important to take into
account that $^{14}$C has an unusually large log-ft value of 9.04. Its
mirror, $^{14}$O, has a value of 3.4892. Thus, while the difference in
ground state energies is only 80~keV, the wave functions may not be
exact mirrors. The spectra are similar, but the first excited state
energies differ by 920 keV.  As a result, while this indicates that
there may be a difference in energy due to Coulomb effects one cannot
estimate it with any certainty due to the anomalous log-ft value for
$^{14}$C. No shell model wave function has been able to reproduce that
large value~\cite{Ka95}.

In this investigation, these differences are taken into account by
(small) variations in the OPP. (See Section~\ref{pv}.)


\subsection{The charge distribution and electromagnetic properties}
\label{betas}

For any nuclear charge distribution, electric multipole operators are
defined in the space-fixed frame by
\begin{equation}
T_{\lambda \mu} = \int \rho_{ch}({\bf r})\ r^\lambda 
Y_{\lambda \mu}^\star(\theta, \phi) \ d{\bf r} .
\label{EMmultop}
\end{equation}
Here $\mu\hbar$ is the angular momentum projection on the space-fixed
z-axis.  

We suppose that the nucleus is like  
an incompressible liquid drop whose surface, $R(\theta, \phi)$, 
can be deformed.  Expanding that surface to first order gives
\begin{equation}
R(\theta \phi)  = R_0 \left[ 1 +
\sum_{\lambda \mu} \alpha_{\lambda \mu}^\star
Y_{\lambda \mu}(\theta \phi) \right] \ .
\label{Rdef}
\end{equation}
Then any function with that surface can also be expanded as
\begin{equation}
F({\bf r}) = F(r) - R_0 \frac{dF}{dr}
\sum_{\lambda \mu} \alpha_{\lambda \mu}^\star Y_{\lambda \mu}(\theta \phi)  ,
\label{Vib-fst}
\end{equation}
and, in particular, the nuclear charge distribution  as 
\begin{equation}
\rho_{ch}({\bf r}) = \rho_0\ \rho_{ch}(r) - 
\rho_0 R_0 \left(\frac{d \rho_{ch}(r)}{d r}\right) 
\sum_{l m} \alpha^\star_{l m} Y_{l m}(\Omega) \ .
\label{chgden}
\end{equation}
Here $\rho_0$ is the central charge density value,
$\rho_0~=~Ze/\left[4\pi \int \rho_{ch}(r) r^2 dr \right]$.

Substituting  Eq.~(\ref{chgden}) in Eq.~(\ref{EMmultop}) gives
\begin{align}
T_{\lambda \mu} &= \int \rho_{ch}({\bf r})\ r^\lambda 
Y_{\lambda \mu}^\star(\theta \phi) \ d{\bf r} \nonumber\\ 
&=  - \rho_0\ R_0 \int_0^\infty r^{\lambda + 2}
\left(\frac{d \rho_{ch}(r)}{dr}\right) \ dr 
\ {\alpha}_{\lambda \mu}^\star \ .
\label{EMmultws}
\end{align}
Quantisation with the collective vibration model is then made using
the transformation
\begin{equation}
\alpha_{\lambda \mu}^* \rightarrow 
\beta_\lambda \frac{1}{\sqrt{(2\lambda + 1)}} 
\left\{  b^\dagger_{\lambda \mu} + (-)^\mu b_{\lambda -\mu}
\right\} ;
\end{equation}
$b^\dagger_{\lambda \mu}$ and $b_{\lambda -\mu}$ are phonon creation
and annihilation operators and $\beta_\lambda$ are coupling strengths.

First order expansions suffice for transitions between pure vibration
model states; the ground as the vacuum ($| 0, 0 >$), and the $2^+_1$ and
$3^-_1$ ones being a single quadrupole and single octupole phonon
excitation upon that vacuum, $ b^\dagger_{2\mu} | 0, 0 >$ and $
b^\dagger_{3\nu} | 0, 0 >$ respectively.  Electromagnetic transitions
between the ground state and the single phonon excited states have
matrix elements of the form
\begin{align}
\left<J_f M_f\left| \alpha_{J_f M_f}^\star \right| 0,0 \right>& \nonumber\\ 
& \hspace{-1cm} = \left< 0, 0 \left| b_{J_f M_f}\; \beta_\lambda 
\frac{1}{\sqrt{2J_f +1}} 
b^\dagger_{J_f M_f}\; \right| 0, 0\right> \nonumber\\
& \hspace{-1cm} = \frac{1}{\sqrt{2J_f+1}} \beta_{J_f}\ . 
\end{align}
with which the electromagnetic transition probabilities are
\begin{equation}
B(E\lambda) = \sum_{M_f} 
\left| \left<J_f M_f\left| T_{\lambda \mu} \right| 0 0 \right> 
\right|^2 =  
\left| \left<J_f\left|\left| T_{\lambda} \right|\right| 0 \right> 
\right|^2 .
\end{equation}
have $\lambda = J_f$.  For the finite distribution of charge, these
transition probabilities are given by
\begin{equation}
B(E\lambda)\uparrow = 
\frac{1}{(2J_f + 1)}  \beta_{J_f}^2\  \rho_0^2\ R_0^2\ 
\left[ \int_0^\infty r^{\lambda + 2} \left(\frac{d\rho_{ch}(r)}{dr} 
\right) \right]^2 .
\end{equation}

We use this pure vibration model to describe the states of ${}^{14}$O
in MCAS evaluations of the spectra of ${}^{15}$F treated as the
$p$+${}^{14}$O cluster, and of low-energy scattering of ${}^{14}$O ions
from hydrogen.  Quadrupole and octupole coupling constants are
involved in defining the matrix of interaction potentials to be used
and for these we usually seek guidance from electromagnetic properties
of the `target'. The relevant $B(E2)\uparrow$ and $B(E3)\uparrow$
values in ${}^{14}$O are as yet unknown, while those values for the
transitions in ${}^{14}$C are uncertain, though from that $B(E2)$
value, Raman~\cite{Ra01} gives an adopted value of $\beta_2 = $0.36
(the sign being ambiguous since $B(E2)$ depends on $\beta_2^2$).
However, we assume that both the $E2$ and $E3$ transitions in
${}^{14}$O would be similar to those in ${}^{16}$O, namely $\sim$40
e$^2$-fm$^4$~\cite{Ra01} and $\sim(1300-1500)$
e$^2$-fm$^6$~\cite{Sp89} respectively.

We have used a 3pF model for the charge distribution in ${}^{14}$O, {\it viz.}
\begin{equation}
\rho_{ch}(r) = \frac{1 + w_c \left(\frac{r^2}{R_c^2}\right)}
{1 + \exp{\left(\frac{r-R_c}{a_c}\right)} } .
\label{3pF}
\end{equation}
As reported in Ref.~\cite{Vr87}, electron scattering form factors,
when used to specify a 3pF charge distribution for ${}^{16}$O, set the
parameter values as $R_c, a_c, w_c$ = 2.608~fm, 0.52~fm, $-$0.051. We
presuppose that the charge distribution in ${}^{14}$O would be
slightly more diffuse and have used the set, $R_c, a_c, w_c$ =
2.59~fm, 0.6~fm, $-$0.051.  With that distribution, the
$B(E2)\uparrow$ with $\beta_2 = $0.36 is 45.6~e$^2$-fm$^4$;
cf. 40.6~e$^2$-fm$^4$ adopted for the transition in ${}^{16}$O.  The
$B(E3)\uparrow$ found using $\beta_3 = 0.48$ is 1323 e$^2$ fm$^6$
which compares with the adopted value of 1300~e$^2$~fm$^6$ for
${}^{16}$O assessed from electron scattering data.  For a full
description of how the 3pF charge distribution is implemented in MCAS,
see Refs.~\cite{Fr15,Fr16}.

\subsection{Parameter values for the nuclear interaction}
\label{pv}

A vibration collective model has been used to specify the matrices of
interaction potentials with the clusters, ${}^{15}$C ($n$+${}^{14}$C)
and ${}^{15}$F ($p$+${}^{14}$O) as has been used
recently~\cite{Sv17}.  The coupled-channel interaction matrices were
formed using the five states in ${}^{14}$C and ${}^{14}$O as 
discussed above. They are listed again in Table~\ref{mass-14-sts} in
which the strengths of the OPP terms required for each are given.  The
OPP scheme is one that allows for Pauli blocking or hindrance of the
added nucleon to the core nucleus in forming the relevant compound
nuclear system.  The OPP strengths listed in Table~\ref{mass-14-sts}
are those that lead to good results for the low excitation spectra for
${}^{15}$C and ${}^{15}$F.  Those values, shown in brackets in the
table, effect fine tuning of the energies of the ${}^{15}$F levels,
notably of the $\frac{1}{2}^-$ state.  This may be a reflection of the
differences between the spectra of the core nuclei, ${}^{14}$C and
${}^{14}$O.

\begin{table}\centering
\setlength\extrarowheight{2.5pt}
\caption{\label{mass-14-sts} Parameter values defining the $n+^{14}$C
  and $p+^{14}$O interaction. $\lambda$ are
  blocking strengths of occupied single nucleon orbits, in
  MeV. Additionally, all states involve a $\lambda$ strength
  of $10^6$ for the $1s_{\frac{1}{2}}$ orbit. The numbers in brackets
  are the OPP values required to give the best representation of the
  three low excitation states in ${}^{14}$F.  Lengths are in fm.}

\begin{supertabular}{>{\centering}p{28mm} >{\centering}p{26mm} 
p{26mm}<{\centering} }
\hline
\hline
                  & Odd parity & Even parity\\
\hline
$V_0$ \ (MeV)       & -48.16  & -43.16\\
$V_{l l}$ \ (MeV)    &  0.475  &  0.475\\
$V_{l s}$ \ (MeV)    &  7.0    &  7.0\\
$V_{ss}$ \ (MeV)     &  0.0    &  0.0\\
\end{supertabular}

\begin{supertabular}{>{\centering}p{13mm} p{10mm}<{\centering} 
>{\centering}p{14mm} >{\centering}p{10mm} >{\centering}p{10mm} 
>{\centering}p{10mm} p{8mm}<{\centering} }
\hline
\hline\\[-1.9ex]
$R_0$\,(fm) & $a$\,(fm) & $R_c$\,(fm) & $a_c$\,(fm) & $w_c$ & $\beta_2$ 
& $\beta_3$\\
3.083 & 0.63 & 2.59 & 0.6 & $-$0.051 & $-$0.36 & $-$0.48 \\
\end{supertabular}

\begin{supertabular}{>{\centering}p{14mm} | p{14mm}<{\centering} 
>{\centering}p{14mm} | >{\centering}p{18mm} p{18mm}<{\centering} }
\hline
\hline\\[-1.9ex]
$J^\pi$ & $E {}^{14}$C & $E {}^{14}$O & $\lambda_{p_\frac{3}{2}}$ 
& $\lambda_{p_\frac{1}{2}}$\\
\hline
$0_1^+$ & 0.00  & 0.00 & 10$^6$ & 9.9 (9.42) \\
$1_1^-$      & 6.09  & 5.17 & 16.0   & \ 5.25 \\
$0_2^+$    & 6.59  & 5.92 & 10$^6$ & 4.2 \\
$3_1^-$      & 6.73  & 6.27 & 16.0   & 4.3 (11.75)  \\
$2_1^+$      & 7.01  & 6.59 & 10$^6$ & \ 2.5 (2.9)  \\
\hline
\hline
\end{supertabular}

\end{table}

The interaction potential strengths required in the MCAS calculations
for the the nucleon - mass-14 clusters were $V_0 = −43.16, V_{ll} =
0.475$, and $V_{ls} = 7.0$ MeV.  The parameter values of the nuclear
interaction geometry used are, $R_0 = 3.063$ fm, $a_0 = 0.62$ fm., and
the deformation parameter values used are $\beta_2 = -0.36$ and
$\beta_3 = -0.48$.  The calculations of the ${}^{15}$F
($p$+${}^{14}$O) system required addition of Coulomb interactions, and
those were derived assuming that ${}^{14}$O had the 3pF charge
distribution given in Eq.~(\ref{3pF}).


\section{Energy levels of {\rm $^{15}$C} and {\rm $^{15}$F}}
\label{spectra}

The spectra of ${}^{15}$C and of ${}^{15}$F found using MCAS are
compared with the experimental values graphically in Figs.~\ref{spect-15C} 
and \ref{spect-15F} respectively. They are discussed in the following
subsection.  The two lowest states in
${}^{15}$C are subthreshold to neutron break-up but all other states
are resonances.  Then in subsection~\ref{EandG}, the centroid
energies and widths, are listed in Tables~\ref{C15-spectrum} and
\ref{F15-spectrum} respectively.

\subsection{Energy level diagrams}

\begin{figure}[h]
\scalebox{0.53}{\includegraphics*{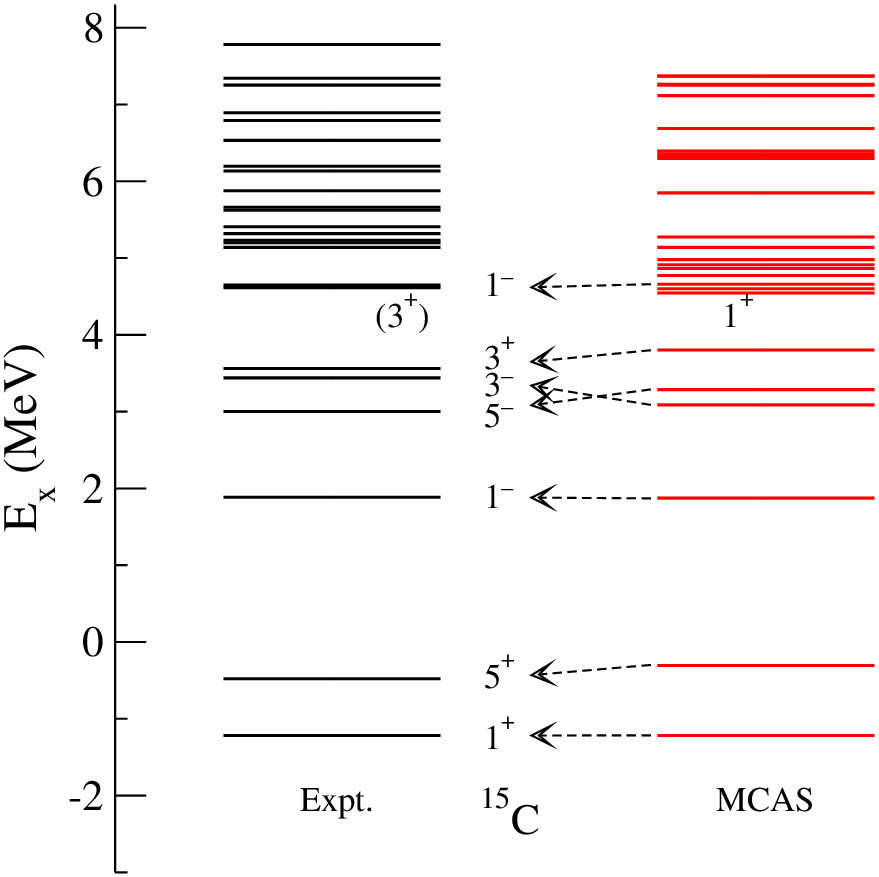}}
\caption{\label{spect-15C}
Spectra of ${}^{15}$C in relation to the $n$+${}^{14}$C threshold.
The states are classified by twice their spin and their parity.
}
\end{figure}

In Fig.~\ref{spect-15C}, the known low-energy spectrum of
${}^{15}$C (to $\sim$8 MeV excitation) is shown in the
column identified by `Expt.'. The lowest eight states have known 
spin-parities. 
They are compared with the spectral results obtained using MCAS with
the vibration model describing the interactions of a neutron with the
five states of the core nucleus, ${}^{14}$C, shown in
Fig.~\ref{Mass-14-lev}. All states other than the lowest two are
resonances and can decay by neutron emission. The lowest six known
states (to $\sim$5 MeV excitation in ${}^{15}$C) are well matched by
the MCAS results save that the order of the close lying
$\frac{5}{2}^-$ and $\frac{3}{2}^-$ resonances is interchanged.  The
energy of the ground state lies 1.217 MeV below the $n$+${}^{14}$C
threshold in good agreement with the experimental value of 1.218 MeV.

\begin{figure}[h]
\scalebox{0.53}{\includegraphics*{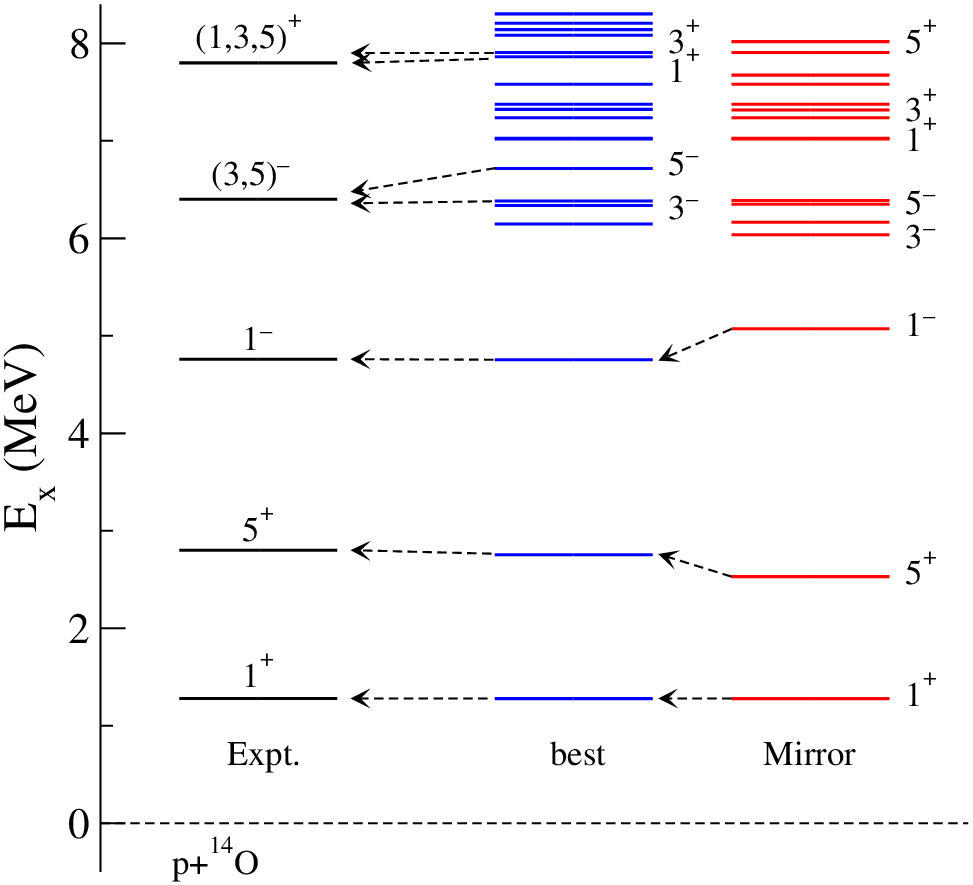}}
\caption{\label{spect-15F}
Spectra of ${}^{15}$F in relation to the $p$+${}^{14}$O threshold.
The states are classified by twice their spin and their parity.
}
\end{figure}

As evident in Fig.~\ref{spect-15F}, little is known of the spectrum of
${}^{15}$F, but the first three resonances have established
spin-parity assignments consistent with the lowest three states in the
mirror, ${}^{15}$C.  Five states of ${}^{14}$O, the mirrors of those
in ${}^{14}$C, were used in the MCAS evaluations for ${}^{15}$F. The
known values of the excitation energies (four being resonance centroid
energies) and the widths of those four resonances, were taken into
account in the coupled-channel calculations.  The relevant Hamiltonian
initially, was taken as that deemed best in giving the spectrum of
${}^{15}$C from the $n$+${}^{14}$C cluster evaluation, with the
addition of Coulomb interactions formed using the 3pF model of the
charge distribution in ${}^{14}$O.  The results of that initial
evaluation are those shown in Fig.~\ref{spect-15F} and labelled
therein by `mirror'.  There is reasonable comparison with the known
spectrum (`Expt.'). Small adjustments made by variation of the
$\lambda_{p_{\frac{1}{2}}}$ values in the OPP set give the results
identified as `best'.  Importantly both evaluations lead to the ground
state resonance lying at 1.279 MeV in the $p$+${}^{14}$O center of
mass.  Clearly there are many more states predicted to lie in the
spectrum above the three, well established, resonances.

\subsection{Tabulated level energies and widths}
\label{EandG}

The two lowest states in ${}^{15}$C are subthreshold to neutron
break-up but all other states are resonances.  The widths determined
by the MCAS evaluations are solely those for nucleon break-up of the
mass-15 systems. With ${}^{15}$F, the lowest two resonances only decay
by proton emission and the measured and calculated widths can be
compared.  We list the values for ${}^{15}$C that are given in
Ref.~\cite{Aj91} but for the three lowest states in ${}^{15}$F we have
used the values assessed in a recent article~\cite{Gr16}. (See
Ref.~\cite{Ke16} for discussion of other measured results.)

\begin{table}[h]
\setlength\extrarowheight{3.5pt}
\caption{\label{C15-spectrum} Spectra of ${}^{15}$C.  The experimental
  values, Expt., are compared with the MCAS results found using the
  vibration model.  All resonance centroid $E_r$ and (full) width
  $\Gamma$ values are in MeV.}
\begin{supertabular}{>{\centering}p{14mm} p{10mm}<{\centering} 
>{\centering}p{14mm} | >{\centering}p{10mm} >{\centering}p{14mm}
p{10mm}<{\centering} }
\hline
\hline
 & Expt. & &  & MCAS & \\
$J^\pi$ & $E_r$ & $\Gamma$ & $J^\pi$ & $E_r$ & $\Gamma$ \\
\hline
\ $\frac{1}{2}^{+}$ & {\bf $-$1.218} & \ ---\  &  $\frac{1}{2}^{+}$ & 
 {\bf $-$1.217} & \ ---\  \\

\ $\frac{5}{2}^{+}$ & {\bf $-$0.478} & \ ---\  & $\frac{5}{2}^{+}$ &  
 {\bf $-$0.3056} & \ ---\  \\

$\frac{1}{2}^{\_}$ & \ {\bf 1.885} & $<$ 0.040 & $\frac{1}{2}^{\_}$  & 
\ {\bf 1.874} & 0.019 \\

$\frac{5}{2}^{\_}$ & \ {\bf 3.002} & $<$ 0.014 & $\frac{5}{2}^{\_}$  &  
\ {\bf 3.287} & 0.003 \\

$\frac{3}{2}^{\_}$ & \ {\bf 3.439} & & $\frac{3}{2}^{\_}$ & 
\ {\bf 3.088} & 0.028 \\

\ $\frac{3}{2}^+$ & \ {\bf 3.562} & 1.74 & $\frac{3}{2}^+$ &  
\ {\bf 3.802} & 3.16 \\

 & & & $\frac{1}{2}^+$ & \ 4.545 & 0.218 \\

 & & & $\frac{3}{2}^-$ & \ 4.600 & 0.009 \\

\ ($\frac{3}{2}^+$) & \ {\bf 4.615} & 0.064 &  $\frac{3}{2}^{+}$  & 
\ {\bf 4.980} & 0.337 \\

\ $\frac{1}{2}^{\_}$ &  \ {\bf 4.648} & & $\frac{1}{2}^{\_}$  & 
\ {\bf 4.648} & 0.006 \\

\hline

($\frac{5}{2},\frac{7}{2}, \frac{9}{2}^+$)
& {\bf \ 5.14\ }  & $<$ 0.02 &  $\frac{5}{2}^{+}$  & {\bf \ 5.140} & 0.297 \\

 & & &  $\frac{5}{2}^{-}$ & \ 5.275 & 0.009 \\

($\frac{3}{2} \to \frac{7}{2}$)
 & \ 5.2\ \   & $\sim$0.05 &  $\frac{1}{2}^{+}$ & \ 5.849 & $7\time 10^{-5}$\\

& & &  $\frac{7}{2}^{-}$ & \ 6.300 & 0.037 \\

& & &  $\frac{5}{2}^{+}$ & \ 6.232 & 0.032 \\

\hline
\hline
\end{supertabular}
\end{table}

\begin{table}[h]
\setlength\extrarowheight{3.5pt}
\caption{\label{F15-spectrum} Spectra of ${}^{15}$F. The notation is
  as given in Table~\ref{C15-spectrum}.  N.B.: In  Ref.~\cite{Mu10},
  Table I gives widths for their $(\sfrac{5}{2},\sfrac{3}{2}^)-$ state
  as 0.2(2), the text on page 10 indicates that 0.2 MeV is the 
  experimental resolution.}
\begin{supertabular}{>{\centering}p{5mm} p{12mm}<{\centering} 
p{9mm}<{\centering} >{\centering}p{21mm} | >{\centering}p{8mm}
>{\centering}p{12mm} p{9mm}<{\centering} }
\hline
\hline
& & Expt. & &  & MCAS & \\
Ref. & $J^\pi$ & $E_r$ & $\Gamma$ & $J^\pi$ & $E_r$ & $\Gamma$ \\
\hline
\cite{Gr16} &\ $\frac{1}{2}^{+}$ & {\bf 1.270} & 0.376$\pm$0.070 &  $\frac{1}{2}^{+}$ & 
{\bf 1.280} & 0.708 \\
\cite{Gr16} &\ $\frac{5}{2}^{+}$ & {\bf 2.794} & 0.30$\pm$0.010 & $\frac{5}{2}^{+}$ & 
{\bf 2.651} & 0.336 \\
\cite{Gr16} &\ $\frac{1}{2}^{-}$ & {\bf 4.757} & 0.036$\pm$0.014 &  $\frac{1}{2}^{-}$ & 
{\bf 4.755} & 0.106 \\
&& &  & $\frac{3}{2}^{-}$ & 6.148 & 0.286 \\
&& &  & $\frac{1}{2}^{-}$ & 6.336 & 0.509 \\
&& &  & $\frac{3}{2}^{-}$ & {\bf 6.384} & 0.951 \\
\cite{Mu10} &\ ($\frac{3}{2}, \frac{5}{2}^{-}$) & {\bf 6.4 } & $\le 0.2$ &  
 $\frac{5}{2}^{-}$ & {\bf 6.717} & 0.074 \\
&&  &  & $\frac{1}{2}^{+}$ & 7.018 & 0.257 \\
&&  &  &  $\frac{7}{2}^{-}$ & 7.027 & 0.176 \\
&&  &  &  $\frac{5}{2}^{-}$ & 7.323 & 0.145 \\
&&  &  &  $\frac{3}{2}^{+}$ & 7.376 & 0.155 \\
&&  &  & $\frac{1}{2}^{-}$ & 7.580 & 0.062 \\
&&  &  & $\frac{1}{2}^{+}$ & 7.862 & 0.052 \\
\cite{Mu10} &\ ($\frac{3}{2}, \frac{5}{2}^{+}$) & {\bf 7.8\ \ } & 0.4$\pm$0.4 &  
$\frac{3}{2}^{+}$ & {\bf 7.906} & 0.134 \\
&&  &  &  $\frac{7}{2}^{-}$ & 8.084 & 0.436 \\
\hline
\hline
\end{supertabular}
\end{table}

The coupled-channel (nuclear) interaction Hamiltonian and the OPP
accounting for Pauli blocking and/or hindrance in the selected five
states of ${}^{14}$C, were chosen to give an optimal match to the
known lowest eight states in ${}^{15}$C.  To emphasise that, those
with energies within $\sim$300 keV of the data are shown in bold face
type in Table~\ref{C15-spectrum}.  There are many more states
predicted by this collective model evaluation.  Above 5 MeV in the
spectrum listed in Table~\ref{C15-spectrum}, the experimentally known
resonances have ambiguous spin-parity assignments though the richer
evaluated spectra have characteristics consistent with those sets.
The widths of the first two resonances are small and consistent with
observation.

The coupled-channel interaction potentials so found were then used
with MCAS to define a spectrum for ${}^{15}$F.  But, as described
earlier, some essential changes to the input specifications had to be
made. First most states of the mirror core nucleus, ${}^{14}$O, are in
fact resonances themselves and were used as such in the MCAS
evaluations.  The excitation energy centroids of those states differ
slightly from the corresponding ones in ${}^{14}$C.  Then there are
Coulomb interactions to be included with the $p$-${}^{14}$O cluster
evaluations.  To find the best representation of the ${}^{15}$F
spectrum, small adjustments to the $\lambda_{p_{\frac{1}{2}}}$ OPP
values as indicated in Table~\ref{mass-14-sts}, were made.  The
results are given in Table~\ref{F15-spectrum} where they are compared
with the limited known spectral values~\cite{Gr16,Mu10}.  

The three best determined resonances, centroid energies and widths,
are quite well matched by the calculation results as are the other two
higher excitation resonances that have uncertain spin-parities and
widths.  The widths of the resonances found with MCAS link solely to
the states decay by single proton emission, and since the higher lying
resonances in ${}^{15}$F can also decay by a two proton emission
process, the widths given in Ref.~\cite{Mu09} would include effects of
that process of decay.

\section{{\rm $^{14}$O} scattering from Hydrogen at 180$^\circ$}
\label{xsect}

\begin{figure*}[h]
\scalebox{0.8}{\includegraphics*{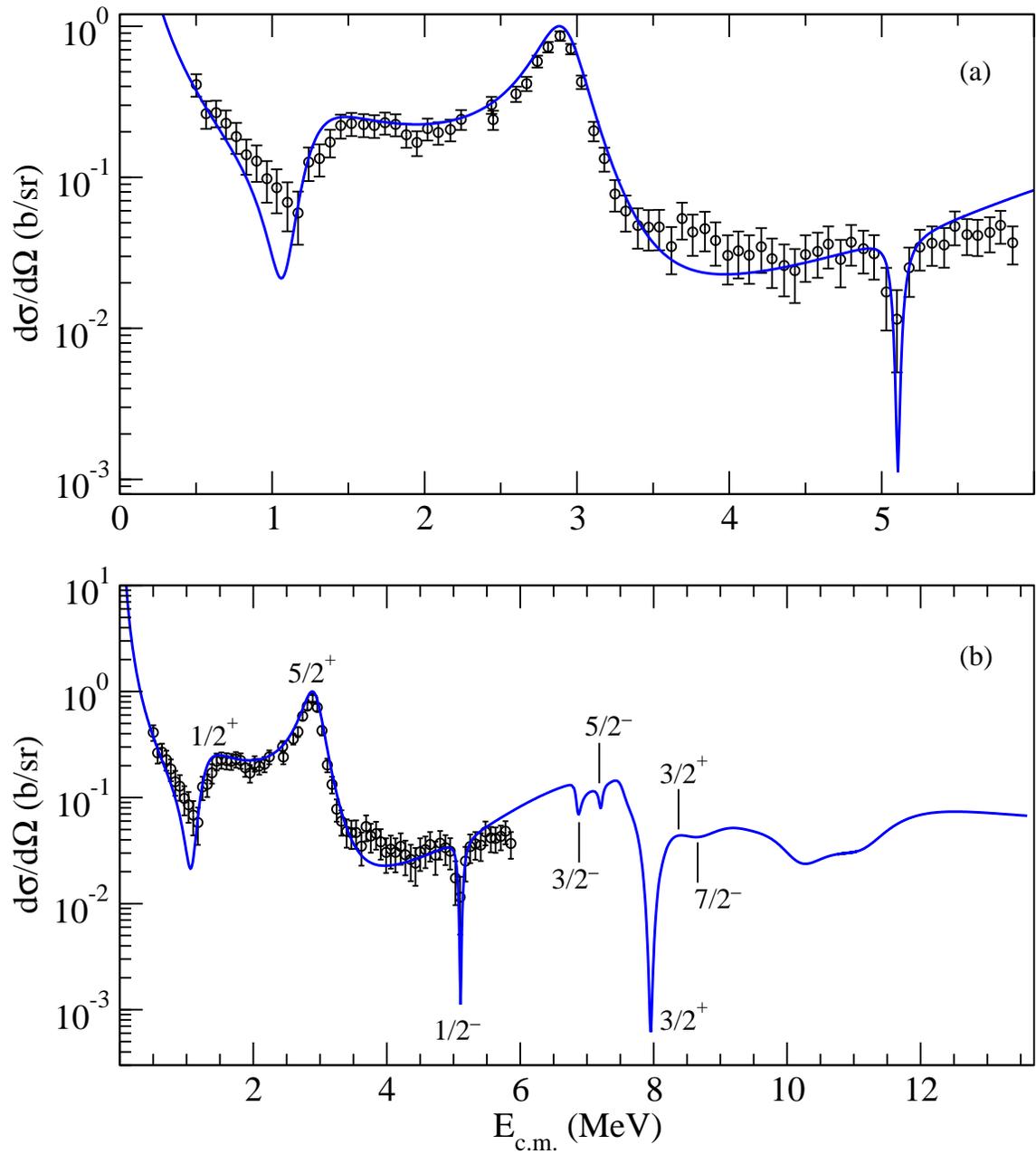}}
\caption{\label{p-14O-180} (a) The 180$^\circ$ cross-section data
  of~\cite{Gr16} compared with MCAS results. The measured data are
  compared with cross sections evaluated using the vibrational model
  to specify the matrix of interaction potentials for a
  proton-${}^{14}$O cluster. Five states of the core as described in
  text were used. (b) The same calculation over a larger energy range,
  predicting resonance features at energies above those measured.}
\end{figure*}

Using five states in the low excitation spectrum of ${}^{14}$O, the
ground ($0^+_1$), the $1^-$ (5.17 MeV), the $0^+_2$ (5.92 MeV), the
$3^-$ (6.27 MeV), and the $2^+$ (6.59 MeV), MCAS calculations gave the
cross sections for $p$-${}^{14}$O scattering at $180^\circ$ that
are compared with data~\cite{Gr16} in Fig.~\ref{p-14O-180}.

Using semilogarithmic graphing emphasizes small structures in both
data and evaluated results. This data clearly indicate three
resonances; in~\cite{Gr16} they are defined as the $\frac{1}{2}^+$
ground state of ${}^{15}$F centered at 1.27 MeV with a width of 0.376
MeV, the $\frac{5}{2}^+$, first excited state, with a centroid and
width of 2.794 and 0.301 MeV, and a $\frac{1}{2}^-$ resonance with
centroid and width of 4.754 and 0.036 MeV respectively.  The
calculated results reproduce those three resonances very well.
Panel~(b) reveals that more structure is predicted for energies in the
region above 6 MeV where we anticipate there exist groups of states of
both parities.  By studying correlations in two proton emission from
${}^{16}$Ne~\cite{Mu10}, two resonance aspects of ${}^{15}$F were
defined in that region having centroids at 6.57 and 7.8 MeV
excitation. But their spin values and widths are uncertain as yet.\\

\section{The mirror pair $^{15}$O ($n$+$^{14}$O) and $^{15}$N ($p$+$^{14}$C)} 
\label{ON}

Being strongly bound, the spectra of the mirror pair, $^{15}$O and
$^{15}$N, have been studied for many decades~\cite{Wy49}. That
of the better known, $^{15}$N, was recently surveyed experimentally
over a range of 15 MeV using the $^{14}$N$(d,p)^{15}$N
reaction~\cite{Me15}, and in the same paper the COSMO shell model
code~\cite{Vo09} was used to successfully calculate these levels, up to
11.5 MeV.  That investigation used an unrestricted $1p-2s1d$ shell
valence space. Mirror states in the less-well-known $^{15}$O spectrum
were then suggested. Another shell model investigation using a lesser
space, the $2s_{1/2}$ and $1d_{5/2}$ shells, soon followed~\cite{Fo16}. 
%

Using MCAS, the nuclear interactions for the $n$+${}^{14}$O and $p$+${}^{14}$C
systems are stronger than those required with the $p$+${}^{14}$O and
$n$+${}^{14}$C calculations. That is evident from the much larger
energies (13.223 and 10.207 MeV) of the relevant nucleon-core nucleus
thresholds above the ground states of ${}^{15}$O and ${}^{15}$N
respectively. That expectation also follows from the numbers of strong
attractive (8), versus those of repulsive (6), two-nucleon
interactions experienced by the extra-core nucleon in the clusters,
${}^{15}$N ($p$+${}^{14}$C) and ${}^{15}$O ($n$+${}^{14}$O).  In the
clusters, ${}^{15}$C ($n$+${}^{14}$C) and ${}^{15}$F ($p$+${}^{14}$O),
in contrast there are 6 strong attractive and 8 repulsive pairings.
Additionally the OPP strengths for the ${}^{15}$N and ${}^{15}$O cases
will differ from those of the ${}^{15}$C and ${}^{15}$F clusters
since, with 6 rather than 8 extra core-like nucleons, the
single-nucleon shell occupancies of those nucleons in the core nuclei
are lesser.

We have used the MCAS approach to optimally find the sub-threshold
levels in ${}^{15}$O treated as the $n$+${}^{14}$O cluster and
especially to find that the ground state lies 13.22 MeV below the
neutron emission threshold. This threshold lies well above those for
emission of a proton (7.30 MeV), an $\alpha$ (10.22 MeV) and a $^3$He
(12.08 MeV).  Thus, while the MCAS calculations lead to 12
sub-threshold (to neutron emission) levels, only the most bound set of
6 are not resonances for emission of the other nuclear particles.
Empirically there are 7 actual sub-threshold bound states in
${}^{15}$O while there are $\sim$ 40 resonant states above those and
below the neutron emission threshold.  On the other hand, the proton
emission threshold in the mirror system, ${}^{15}$N, is the first of
such and lies 10.207 MeV above the ground.  Empirically, there are 17
sub-threshold (bound) states in ${}^{15}$N.

\subsection{Specifics of the ${}^{15}$O and ${}^{15}$N evaluations using MCAS}

The nuclear interaction and the OPP weights to account for Pauli
blocking of single nucleon states was specified by finding as good a
spectrum for ${}^{15}$O ($n$+${}^{14}$O) as possible. In particular,
we sought the ground state of correct spin-parity and energy below the
neutron emission threshold and the first two excited states in the
correct order and with good energy values.  The coupled-channel
Hamiltonian was formed using the five states of the target nucleus
${}^{14}$O as used before (and of ${}^{14}$C in the case of
${}^{15}$N). The geometry, $V_{ls}$, and $V_{ll}$ values were set at
those determined from our MCAS study of the other mass-15 isobars,
${}^{15}$C and ${}^{14}$F.  However, for the reasons discussed above,
the central interaction strength was varied with $-57.0$ MeV found
appropriate.  The parameter values of the OPP used for the two systems
are listed in Table~\ref{15O-param}.

\begin{table}[h]
\setlength\extrarowheight{2.5pt}
\caption{\label{15O-param}
Parameter values defining the $n+^{14}$O
  and $p+^{14}$C interaction interactions. $\lambda$ are
  blocking strengths of occupied single nucleon orbits, in
  MeV.}

\begin{supertabular}{>{\centering}p{28mm} >{\centering}p{26mm} 
p{26mm}<{\centering} }
\hline
\hline
                  & Odd parity & Even parity\\
\hline
$V_0$ \ (MeV)       & $-$57.0  & $-$57.0\\
$V_{l l}$ \ (MeV)    &  0.475  &  0.475\\
$V_{l s}$ \ (MeV)    &  7.0    &  7.0\\
$V_{ss}$ \ (MeV)     &  0.0    &  0.0\\
\end{supertabular}
\begin{supertabular}{>{\centering}p{10mm} p{14mm}<{\centering} 
>{\centering}p{14mm} >{\centering}p{14mm} >{\centering}p{14mm} 
p{10mm}<{\centering} }
\hline
\hline\\[-1.9ex]
& $R_0$\,(fm) & $a$\,(fm) &  $\beta_2$ & $\beta_3$    &\\
& 3.083 & 0.63 & $-$0.36 & $-$0.48 &\\
\end{supertabular}
\begin{supertabular}{>{\centering}p{10mm} | p{12mm}<{\centering} 
p{12mm}<{\centering} |
>{\centering}p{14mm} >{\centering}p{14mm} p{14mm}<{\centering} }
\hline
\hline\\[-1.9ex]
$J^\pi$ & E ${}^{14}$O &  E ${}^{14}$C & $\lambda_{0s_\frac{1}{2}}$ & 
$\lambda_{0p_{\frac{3}{2}}}$ & $\lambda_{0p_{\frac{1}{2}}}$ \\
\hline
$0_1^+$ & (0.00) & (0.00) & 10$^6$ & 17.5 & 2.8\ \\
$1_1^-$ & (5.17) & (6.09) & 10$^6$ & 17.5 & 1.25\\
$0_2^+$ & (5.92) & (6.59) & 10$^6$ & 17.5 & 3.5\ \\
$3_1^-$ & (6.27) & (6.78) & 10$^6$ & 17.5 & 1.6\ \\
$2_1^+$ & (6.59) & (7.01) & 10$^6$ & 17.5 & 1.7\ \\
\hline
\hline
\end{supertabular}
\end{table}

For the ${}^{15}$N calculation Coulomb interactions have been added to
the nuclear ones and the appropriate set of state energies in
${}^{14}$C used.  In this case, the Coulomb interactions were
constrained by using a charge distribution that matches the known
root-mean-square (rms) charge radius.  For ${}^{14}$C that value is
$R^{(c)}_{\rm rms} = 2.56 \pm 0.05$ fm~\cite{Vr87}, defined using a
modified Harmonic Oscillator (MHO) model for the charge distribution
of ${}^{14}$C to analyze electron scattering form factor in the
momentum range 1.04 to 2.16 fm$^{-1}$.

We have used the three parameter Fermi (3pF) model for the charge
distribution.  Sets of parameter values ranging between those
reported~\cite{Vr87} from analyses of electron scattering data from
${}^{12}$C and from ${}^{16}$O were determined by the distributions
having the charge rms radius of 2.56 fm for ${}^{14}$C.  That set of
parameters are listed in Table~\ref{m15-3pF}.
\begin{table}[h]
\setlength\extrarowheight{2.5pt}
\begin{ruledtabular}
\caption{\label{m15-3pF}
Parameter values for a 3pF model of the charge distribution in
${}^{14}$C that give $R^{(c)}_{\rm rms} = 2.56$ fm and the
ground state energies from MCAS calculations of the
$p$+${}^{14}$C system.}
\begin{tabular}{ccccc}
ID & $R_c$ fm. & $a_c$ fm. & $w_c$ & $E_{\rm g.s.}({}^{15}$N)\\
\hline
(a) & 2.355 & 0.5224 & $-$0.08 & $-$10.200\\
(b) & 2.355 & 0.6 & $-$0.149 & $-$10.209\\
(c) & 2.52 & 0.5224 & $-$0.149 & $-$10.206\\
\hline
(d) & 2.355 & 0.5 & $-$0.04 & $-$10.199\\
(e) & 2.355 & 0.54 & $-$0.1 & $-$10.203\\
(f) & 2.355 & 0.64 & $-$0.15 & $-$10.232\\
\hline
(g) & 2.425 & 0.5 & $-$0.06 & $-$10.208\\
(h) & 2.525 & 0.5 & $-$0.09 & $-$10.222\\
(i) & 2.536 & 0.5 & $-$0.1 & $-$10.220
\end{tabular}
\end{ruledtabular}
\end{table}
The first three parameter sets, (a), (b), and (c) in
Table~\ref{m15-3pF} are one parameter variations on the 3pF model
parameters for the adopted charge distribution in
${}^{12}$C~\cite{Vr87} that give $R^{(c)}_{\rm rms} = 2.56$ fm.  The
set, (d), (e), and (f), kept $R_c - 2.344$ fm, varied $a_c$ and
adjusted $w_c$ to find the same $R^{(c)}_{\rm rms}$.  The last set in
Table~\ref{m15-3pF} kept $a_c = 0.5$ fm varied $R_c$ and adjusted
$w_c$ to have the same result.  Thus there are quite diverse sets of
parameters for this model giving the known rms charge radius.  As
shown in Refs.~\cite{Fr15,Fr16}, it is the value of $R^{(c)}_{\rm
  rms}$ that affects the Coulomb potential, with the specific values
of $R_c, a_c$, and $w_c$ leading to that $R^{(c)}_{\rm rms}$ being
only of minor impact on results.
MCAS evaluations using each of these 3pF sets were made and the spectra
found were all very similar.  The last column in Table~\ref{m15-3pF}
lists the value of the ground state energies showing a difference 
of at most 20 keV. 

\subsection{The spectra of ${}^{15}$O and ${}^{15}$N}
\label{results-15O-15N}

\begin{figure*}[h]
\scalebox{0.7}{\includegraphics*{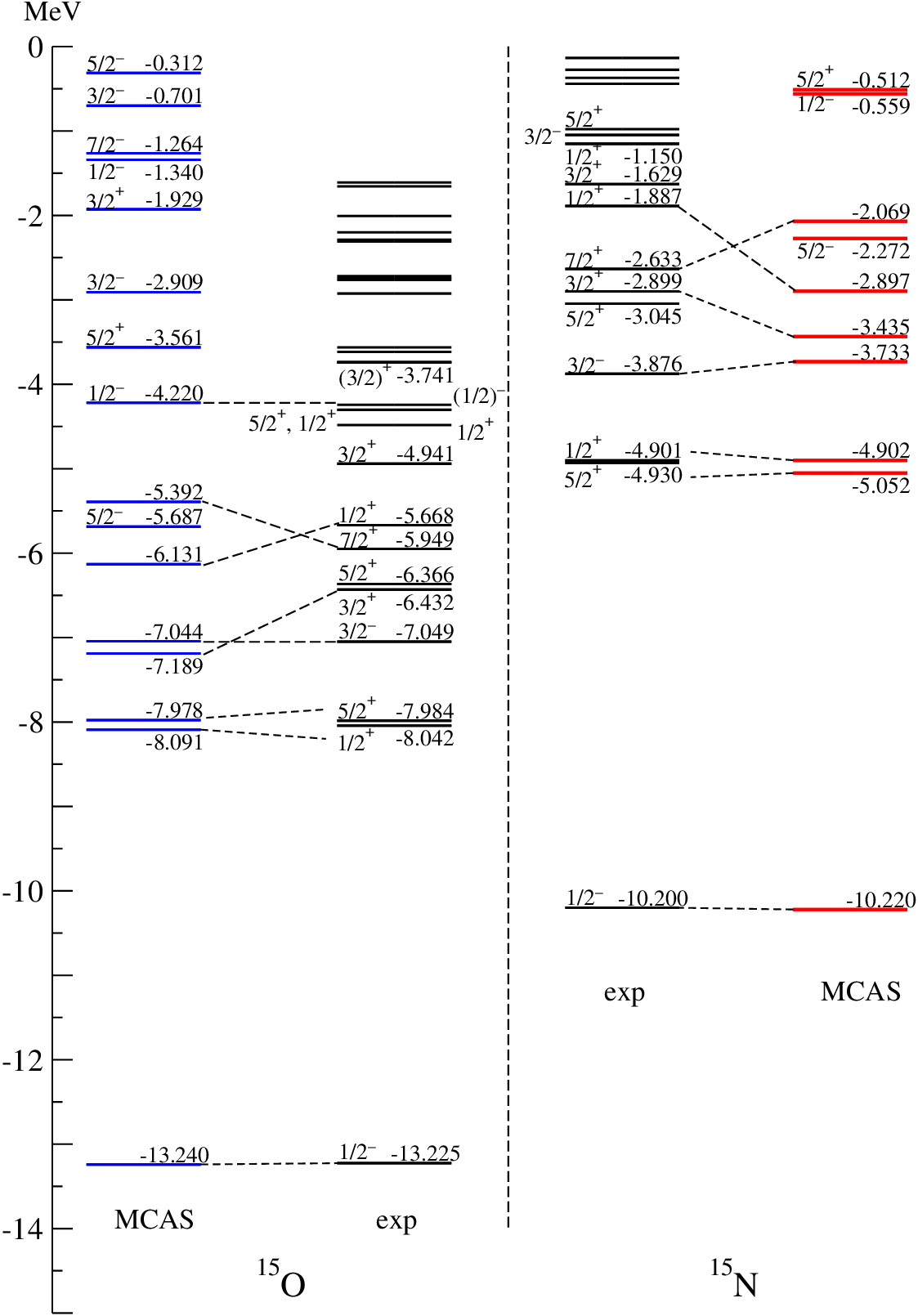}}
\caption{\label{M15-spect}
The sub-threshold spectra of $^{15}$O (left) and of ${}^{15}$N (right)
found using MCAS compared to the experimental values.}
\end{figure*}

The low-excitation spectra of the mirror pair, ${}^{15}$O and ${}^{15}$N,
are depicted in Fig.~\ref{M15-spect}. 
Those for ${}^{15}$O are shown on the left
and those for ${}^{15}$N on the right. The excitation energies are
shown relative to the nucleon separation thresholds 
(13.223 MeV for $n$+${}^{14}$O
and 10.207 MeV for $p$+${}^{14}$C). The calculated spectrum for ${}^{15}$N
displayed  was found using the 3pF model 
parameter set `(h)' in Table~\ref{m15-3pF}.
The known states are given in the columns labelled `exp' were 
taken from ~\cite{Aj91} and the calculated spectra are 
identified by the label `MCAS'.

The results for ${}^{15}$O  closely match  most known
states to 10 MeV excitation though there are many more
levels that lie $\sim 10$ MeV and greater above the ground.
Notably, the ground state is found to within a
keV of its known energy below the neutron emission threshold,
the doublet of states at $\sim -8$ MeV binding are found in the correct 
order, and the next five known states in the spectrum 
have calculated partners with energies within a few hundred
keV of the known values.

With that nuclear interaction, using the appropriate energies of the 
same five target states (in ${}^{14}$C), and the 
3pF charge distribution with the parameters of set `(h)' 
in Table~\ref{m15-3pF}, the single run of MCAS then lead to
the spectrum for ${}^{15}$N that is compared with the known
one in the right side of  Fig.~\ref{M15-spect}. The ground
state was found to be $-10.22$ MeV below the neutron emission
threshold, in good agreement with the known value,
and the low lying spectrum again reasonably matched.
The $\frac{5}{2}^+\vert_1$ state now is more bound than the
$\frac{1}{2}^+\vert_1$ one, as is the case in the experimental spectrum, 
and the splitting of that doublet is larger than observed.
The next two states in the known spectrum of ${}^{15}$N have matching 
partners from the calculation, both lying within a few hundred keV 
of the appropriate energies. Also the known $\frac{7}{2}^+$ state has a 
calculated partner in close agreement but the $\frac{5}{2}^+\vert_2$ 
state is calculated to be at $\sim -0.5$ MeV, not at $\sim -3$ MeV 
in the experimental spectrum.

The known spectra of both mass-15 isobars are much richer than those
we have evaluated but only for reasonably large excitations reflecting
the simplicity of the model chosen to define the coupled-channel
Hamiltonian; with the number of core nuclear states used and use of
the purest of vibration models for the structure and interactions.


\section{Conclusions}
\label{conc}

Mirror symmetry for nuclear interactions was used to study the spectra
of the mass-15 isobars, ${}^{15}$C, ${}^{15}$N, ${}^{15}$O, and ${}^{15}$F.
The MCAS method has been used to evaluate their low energy
(to $\sim 10$ MeV) excitation spectra considering each to be a cluster
of a nucleon with either of the mirrors, ${}^{14}$C and ${}^{14}$O.  
There are  two mirror pairs in these mass-15 isobars,
${}^{15}$O and ${}^{15}$N, and ${}^{15}$C and ${}^{15}$F, which 
are distinct in that the former are well bound with many uniquely
bound states in their spectra, while of the latter pair 
${}^{15}$C is weakly bound with just two sub-threshold
states and its mirror, ${}^{15}$F, lies beyond the proton drip-line.  

In the evaluations, the lowest five states in the core
nuclei, ${}^{14}$C and ${}^{14}$O, were used to form coupled-channel
interactions based upon a collective (vibration) model description of
the core nuclei.  First we sought the spectra of ${}^{15}$C (as the
$n$+${}^{14}$C cluster). With the 
set of parameter values for the Hamiltonian that gave a best match to 
the known spectrum of ${}^{15}$C (to $\sim 6.5$ MeV excitation,
on addition of Coulomb interaction terms, that Hamiltonian lead
to a good match to the known spectrum of ${}^{15}$F.
Coulomb interactions for the
$p$+${}^{14}$O system were formed from a three parameter Fermi 
model for the charge distribution in ${}^{14}$O.  
The same basic nuclear potential, modified only in central well depth and
with OPP strengths reflecting the changes in like-nucleon shell
occupancies in the cores, was used to evaluate the
spectra of the other mass-15 isobar pair ${}^{15}$O and
${}^{15}$N.  With this essentially single potential matrix in the Hamiltonians
very good agreement was obtained for the low-energy spectra
of ${}^{15}$O and ${}^{15}$N.

Finally, as the MCAS procedure produces scattering phase shifts
for $^{14}$O$(p,p)^{14}$O scattering, and in light of recent data,
the elastic scattering cross-section calculation reported in a previous 
letter~\cite{Ca06} has been updated. Very good agreement has been found
between all three known resonance features and the non-resonant scattering
background.  
These calculated results suggest that scattering cross sections when
measured at higher energies (7-9 MeV for example) should reveal 
more structure (resonance states) in the exotic nucleus, ${}^{15}$F.

\section*{Acknowledgments}
 
SK acknowledges support from the National Research Foundation of South Africa.

\bibliography{Mass-15}

\end{document}